# Active high-entropy photocatalyst designed by incorporating alkali metals to achieve $d^0+d^{10}+s^0$ cationic configurations and wide electronegativity mismatch

Jacqueline Hidalgo-Jiménez[1,2], Taner Akbay[3], Tatsumi Ishihara[4] and Kaveh Edalati[1,4,]*

[1] WPI, International Institute for Carbon Neutral Energy Research (WPI-I2CNER), Kyushu University, Fukuoka 819-0395, Japan
[2] Department of Automotive Science, Kyushu University, Fukuoka, Japan
[3] Materials Science and Nanotechnology Engineering, Yeditepe University, Istanbul, Turkey
[4] Mitsui Chemicals, Inc. - Carbon Neutral Research Center (MCI-CNRC), Kyushu University, Fukuoka, Japan

Photocatalytic hydrogen ($H_2$) production and carbon dioxide ($CO_2$) conversion to methane ($CH_4$) are considered promising solutions for reducing $CO_2$ emissions. However, the development of highly active photocatalysts is essential to efficiently drive these reactions without harming the environment. In this study, we introduce a strategy that incorporates elements with both low and high electronegativities into catalysts based on transition metals, thereby enhancing both reactant adsorption and charge transfer. This strategy is implemented in a high-entropy oxide (HEO) by adding cesium, an alkali metal with very low electronegativity, and gallium, a metal with high electronegativity, to transition metals titanium, niobium and tantalum. The resulting oxide, $TiNbTaGaCsO_9$ with a large concentration of oxygen vacancies, exhibits strong light absorption, a low bandgap and a suitable band structure for both hydrogen evolution and $CO_2$ conversion. Compared to HEOs with only $d^0$ or $d^0+d^{10}$ cationic configurations, the synthesized oxide with a wide electronegativity difference and mixed $d^0+d^{10}+s^0$ cationic configurations shows significantly higher activity for both $H_2$ and $CH_4$ production, even without using a cocatalyst. These results demonstrate a design strategy for creating highly active HEOs containing alkali metals by taking advantage of the electronegativity mismatch across the periodic table.

*Corresponding author (E-mail: kaveh.edalati@kyudai.jp; Tel/Fax: +81 92 802 6744)

# 1. Introduction

Carbon dioxide ($CO_2$) emissions are constantly increasing due to growing energy demand. Two potential strategies for lowering the concentration of $CO_2$ in the atmosphere are carbon capture [1] and clean hydrogen ($H_2$) or methane ($CH_4$) production as a replacement for fossil fuels [2]. However, carbon capture technologies need extra steps that involve storing $CO_2$ and its transformation into useful products, and hydrogen production requires developing zero-emission methods, as most of the current hydrogen production processes utilize fossil fuels [3]. Aiming to improve the pathway for $H_2$ generation and $CO_2$ conversion, photocatalysis [4] is drawing attention. In this method, light interaction with the material excites the electrons from the valence band to the conduction band. Subsequently, the photoexcitation generates charge carriers that participate in the required reactions [4,5]. Most of the popular photocatalysts include transition metals with $d^0$ or $d^{10}$ cationic configurations [5,6], such as $TiO_2$ [7,8], $ZrO_2$ [9], $Nb_2O_5$ [10,11] and $Ta_2O_3$ [12] with $d^0$ configurations and ZnO [13,14] with $d^{10}$ configurations. However, there are several limitations in the utilization of these photocatalysts, like low light absorption, weak charge mobility and limited active sites, which result in low efficiency.

Multiple strategies have been studied to enhance the effectiveness of photocatalysts, some including the utilization of dopants [15–17], defect engineering [18,19], high-pressure phases [8], and, most recently, using high-entropy materials. Doping has been proven to be beneficial because the presence of dopants generates defect states, improving the activity [17]. Among various dopants, it has been particularly reported that the addition of alkali metals to perovskites [20], nitrides [21–23] and oxides [24] can improve the light absorbance and diminish electron-hole recombination [21,24]. Motivated by the effectiveness of the doping method, the utilization of high-entropy ceramics, with a minimum of five cations, has gained popularity for photocatalysis since 2020 [25]. The presence of multiple cartons in these high-entropy catalysts not only allows taking advantage of heavy doping but also gives high flexibility in controlling properties through the cocktail effect [25]. In addition, these materials have some other features that make them appropriate for photocatalysis, such as severe strain in the crystal structure [26–28], heterogeneous electric fields inside the material [29] and the presence of defects like oxygen vacancies [28,30].

High-entropy materials have been used for different photocatalytic reactions within the past few years, such as dye degradation [26,31], pollutants removal [32], $CO_2$ conversion [27,28,30,33] and hydrogen production [25,29,31,34,35]. While the concept of high-entropy photocatalysts has opened unlimited possibilities to tune the composition, the design of the composition is still at a very early stage. The first high-entropy photocatalyst only incorporated $d^0$ cations, but the strategy of combining elements with hybrid $d^0$ and $d^{10}$ electronic configurations was shown to improve the photocatalytic activity [33,34]. First-principles calculations suggested that cations with low electronegativity are effective for reactant adsorption, while cations with high electronegativity can improve electron transfer [36]. Despite these preliminary findings, the activity of high-entropy photocatalysts still needs further improvement by the discovery of new design strategies.

This study introduces mixing elements with a wide electronegativity mismatch as a new strategy for designing active photocatalysts. This concept was realized by combining elements with both low and high electronegativities. As shown in the periodic table in Fig. 1 [37], alkali metals with an $s^0$ electronic configuration have low electronegativities, with cesium having the lowest among them. On the other hand, elements with a $d^{10}$ cationic configuration have high electronegativity. Based on this, a new high-entropy oxide (HEO) with an overall $TiNbTaGaCsO_9$ composition and $d^0+d^{10}+s^0$ cationic configurations is designed, which demonstrates high photocatalytic efficiency for hydrogen and methane formation. This study suggests that the strategy of incorporating a wide electronegativity mismatch within $d^0+d^{10}+s^0$ cationic configurations provides a novel approach for designing active photocatalysts.

Fig. 1. Periodic table of elements and respective electronegativity of elements.

## 2. Experimental Procedures
### 2.1. Synthesis

Four commercial oxides, TiO$_2$ anatase powder (Sigma Aldrich, 99.8%), Nb$_2$O$_5$ (Kojundo, 99%), CsTaO$_3$ (Mitsuwa´s Pure Chemicals, 99.9%) and Ga$_2$O$_3$ (Kojundo, 99.99 %) were hand-mixed for 30 min using a mortar under acetone to achieve a good mixture of the powders with a general composition of TiNbTaGaCsO$_9$. Subsequently, 406 mg of powder mixtures were compressed into a disc with a 1 cm diameter and 0.08 cm height. The compacted discs were mechanically mixed using high-pressure torsion (HPT) [38] under 6 GPa pressure and 1 rpm speed for 3 rotations at atmospheric temperature. The obtained sample was hand-crushed one more time and subjected to the same process to ensure the nanometric mixture of the initial powders. The final disc was crushed one more time, and calcinated in a furnace at 1373 K for 24 h with the purpose of mixing the cations at the atomic scale. The oxide, after calcination, was processed one more time using HPT and calcination to improve the homogeneity of elements at the atomic level.

### 2.2. Characterization

The crystal structure transformation during the synthesis was evaluated using X-ray diffraction (XRD) with the Cu Kα light source. Moreover, the vibrational modes of the final high-entropy sample were analyzed by Raman spectroscopy employing laser light (wavelength: 532 nm). The oxidation state of all cations was evaluated by X-ray photoelectron spectroscopy (XPS) with an Al Kα light. The microstructure and homogeneity of the composition were evaluated by scanning electron microscopy (SEM) at 15 keV, employing an energy dispersive X-ray spectrometer (EDS). A combustion-based oxygen, nitrogen and hydrogen (OHN) analyzer (HORIBA EMGA-930, Japan) with two non-dispersive infrared analyzers was used to examine the content of oxygen. Electron spin resonance spectroscopy (ESR) with a 9.4688-GHz microwave was implemented to confirm the presence of oxygen vacancies with unpaired electrons. The absorbance of light by the catalyst was tested by a UV-vis spectrometer, followed by a bandgap estimation via the Kubelka-Munk theory. UV photoelectron spectroscopy (UPS) using a helium UV light source was employed for examining the valence band maximum position with a bias of -5 V.

## 2.3. Photocatalysis

The activity of $TiNbTaGaCsO_9$ was tested using two different photocatalytic reactions of $H_2$ generation from water splitting and $CH_4$ formation from $CO_2$ conversion. For hydrogen production, three different solutions were tested by irradiation: (i) 50 mg catalyst and 27 cm$^3$ $H_2O$, (ii) 50 mg catalyst, 27 cm$^3$ $H_2O$ and 3 cm$^3$ methanol as a hole scavenger, and (iii) 50 mg catalyst, 27 cm$^3$ $H_2O$, 3 cm$^3$ methanol and 0.25 cm$^3$ 0.01 M $H_2PtCl_6.6H_2O$ as a platinum cocatalyst source. The hydrogen production tests were carried out in a 146 cm$^3$ reactor under a full arc of a 300 W xenon lamp with 1.5 W cm$^{-2}$ intensity. The gas products were evaluated by a gas chromatograph (GC), Shimadzu GF-8A. The test was carried out in a single cycle of 180 min, in which hydrogen content was quantified every 30 min. A four-cycle test was also conducted to examine the reusability of the catalyst. Two blank tests, with the catalyst and without illumination, and without the catalyst and under illumination, confirmed the absence of any hydrogen amounts.

For the $CO_2$ conversion, a single condition was tested using a solution containing 100 mg of the catalyst, 4.2 g of $NaHCO_3$ and 500 cm$^3$ of high-purity $H_2O$. The solution was poured into a reactor made of quartz with a 858 cm$^3$ volume, which was connected to two pipes: one for continuous injection of $CO_2$ with a rate of 30 cm$^3$ min$^{-1}$, and the second as an output coupled to two GCs. One GC (Shimadzu GF-8A) was used to quantify the $H_2$ concentration, and another one (GL Science GC-4000 Plus) to quantify the CO and $CH_4$ concentration. A mechanizer (GL Science MT 221) was used before the second GC to improve the detection limits. As a light source, a 400 W high-pressure mercury lamp was placed in the inner area of the photoreactor. The light intensity on the reactants was 0.23 W cm$^{-2}$. The temperature and homogeneity of the suspension were controlled using a water chiller and a magnetic stirrer, respectively. Before turning the light on, the solution was kept for 1 h under constant gas flow. Moreover, a blank test was carried out with illumination but without the addition of the catalyst to confirm the non-existence of any reaction products.

## 3. Results
### 3.1. Crystal structure and microstructure

XRD was used to analyze the transformation of crystal structures in the synthesis procedure, as illustrated in Fig. 2(a). The starting powders exhibit the peaks of the corresponding initial oxides. Once the sample is processed by HPT, the material exhibits significant peak broadening due to the reduction of crystallite sizes to the nanometer scale [39]. The utilization of HPT has proven to not only reduce the crystallite size but also improve the mixture of powders at the nanoscale [34,40]. After the first calcination, most of the peaks related to the initial oxides disappear and new peaks appear, although some small peaks still match with the initial oxides. However, after the second cycle of HPT and calcination, all these small peaks disappear, which is an indication of the transformation of the initial materials to a new oxide. It should be noted that the HPT method was used to enhance the mixture of the oxides at the nanometric scale, promoting the solid-state reactions during the next calcination step. Although the HPT method can affect the microstructure and defects in ceramics [38], these effects disappear when the materials are calcinated at high temperatures.

Fig. 2(b) shows a closer look at the XRD pattern obtained for the resulting high-entropy compound TiNbTaGaCsO$_9$. The Rietveld analysis showed that the material consists of two phases: 59 wt% hexagonal with the $P\bar{3}m1$ space group and another 41 wt% tetragonal with the $I4/m$ space group (refinement parameters: $R_{WP}$ = 12.34%, $R_P$ = 8.75%, $R_{exp}$ = 8.28% and $s$ = 5.456). The slightly low quality of refinement should be due to the uncertainty in the position of cations in two lattices and perhaps some minor amounts of unreacted phases. Table 1 summarizes the results of Rietveld refinement about space groups, lattice parameters, phase fractions and crystallite sizes, confirming the dual-phase nature of the material with small crystallite sizes.

The Raman spectra are presented in Fig. 2(e) for three different random positions of the sample. About 12 different peaks (117, 208, 241, 367, 397, 506, 569, 631, 822, 860, 895 and 953 cm$^{-1}$) are observed, while the shape of the peaks is similar at different positions. The similarity of peaks at various positions is an indication of the smaller size of the two phases compared to the probe size of Raman spectroscopy. Moreover, this similarity suggests that the Raman peaks originate from both phases, and the distribution of the phases at the micrometer level is uniform in the material This similarity in Raman spectra at different positions was also observed in other dual-phase HEOs reported before, such as TiZrHfNbTaO$_{11}$ (d$^0$ cationic configuration with orthorhombic and monoclinic phases) and TiZrNbTaGaO$_{10.5}$ (d$^0$+d$^{10}$ cationic configuration with orthorhombic and monoclinic phases) [25,30,34].

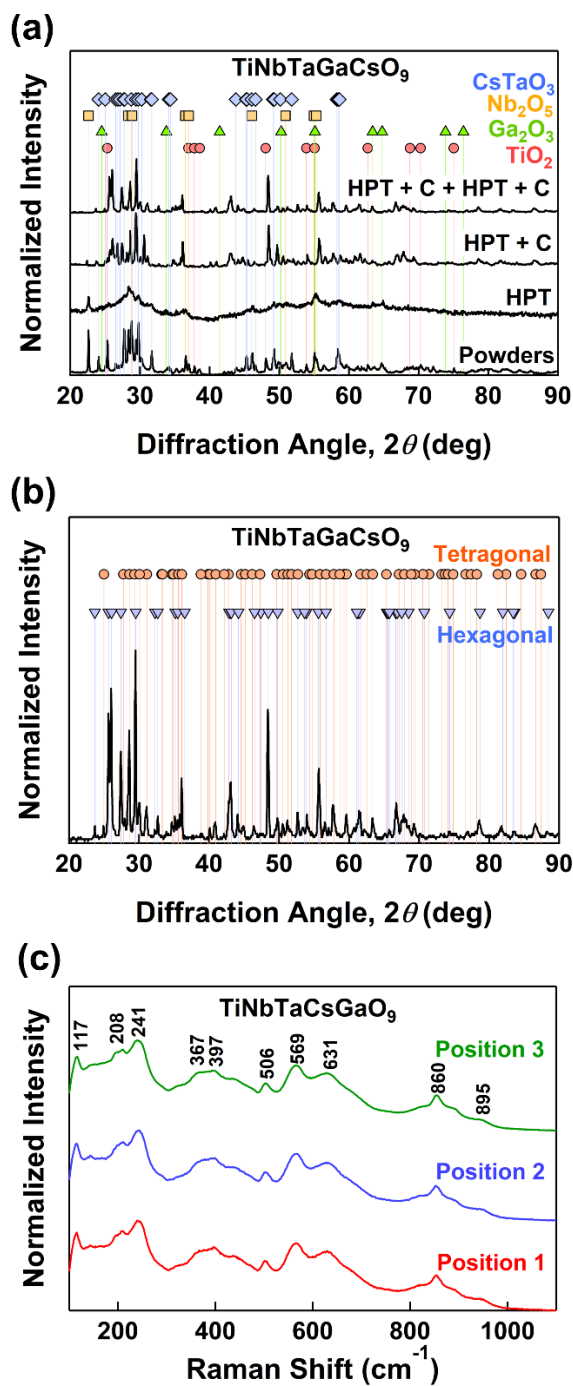

Fig. 2. Synthesis of high-entropy oxide. (a) XRD profiles of initial powder (Powders), HPT-treated sample (HPT), calcinated sample (HPT + C) and sample after repeating HPT and calcination (HPT + C + HPT + C). (b) XRD pattern obtained for final product TiNbTaGaCsO$_9$ with hexagonal and tetragonal phases. (c) Raman spectra achieved at three random positions for final product TiNbTaGaCsO$_9$.

Table 1. Space groups, lattice parameters, phase fractions and crystallite sizes of high-entropy oxide TiNbTaGaCsO$_9$ with hexagonal and tetragonal phases.

|  | Hexagonal | Tetragonal |
|---|---|---|
| Space group | $P\bar{3}m$ | $I4/m$ |
| $a$ (Å) | 7.513 ± 0.004 | 18.181 ± 0.004 |
| $b$ (Å) | 7.513 ± 0.004 | 18.181 ± 0.004 |
| $c$ (Å) | 8.221 ± 0.01 | 3.011 ± 0.001 |
| $\alpha$ (°) | 90 | 90 |
| $\beta$ (°) | 90 | 90 |
| $\gamma$ (°) | 120 | 90 |
| Phase fraction (wt%) | 59 ± 1 | 41 ± 1 |
| Crystallite size (Å) | 495 ± 60 | 239 ± 17 |

SEM-EDS was used to ensure the dispersion of all the elements within the material, as illustrated in Fig. 3. The general composition obtained by EDS is 6.6 ± 0.1 at% titanium, 7.9 ± 0.1 at% niobium, 7.5 ± 0.1 at% tantalum, 7.7 ± 0.1 at% gallium, 6.5± 0.1 at% cesium and 63.8 ± 0.1 at% oxygen, sensibly agrees with the nominal composition of the oxide. The analysis of EDS mappings confirms that TiNbTaGaCsO$_9$ is a dual-phase material, with one of the phases being Ti-Ga-poor and another being Ti-Ga-rich. The composition of the Ti-Ga-poor phase is determined as 2.3 ± 0.1 at% titanium, 11.0 ± 0.9 at% niobium, 9.8 ± 0.8 at% tantalum, 0.8± 0.1 at% gallium, 7.9 ± 0.1 at% cesium and 68.2 ± 2.4 at% oxygen with a configurational entropy of 1.36$R$ ($R$: gas constant), and the composition of the Ti-Ga-rich phase is determined as 14.3 ± 3.3 at% titanium, 2.2 ± 0.5 at% niobium, 2.6 ± 0.5 at% tantalum, 6.3 ± 3.6 at% gallium, 0.8 ± 0.3 at% cesium and 73.8 ± 2.7 at% oxygen with a configurational entropy of 1.22$R$. Since the quantification of light elements such as oxygen using EDS is not so precise, a combustion-based OHN analyzer was used to quantify the amount of oxygen, indicating 56.7 at% oxygen. It appears that EDS overestimates the amount of oxygen due to its limits in analyzing light elements, and OHN analysis underestimates the amount of oxygen due to the presence of hard-to-combust refractory elements. Nevertheless, the calculated values for the entropy (1.36$R$ and 1.22$R$) indicate that this material can be classified either as a composite of two medium-entropy oxides (entropy: 1$R$-1.5$R$) or as a dual-phase HEO (overall entropy: >1.5$R$) [41,42]. In addition to compositional analysis, SEM micrographs suggest that the mean size of particles is 35 μm, corresponding to a surface area of 0.03 m$^2$ g$^{-1}$. The low surface area is because of the synthesis procedure, which involves using high pressure through HPT and high temperature through the calcination treatment [39].

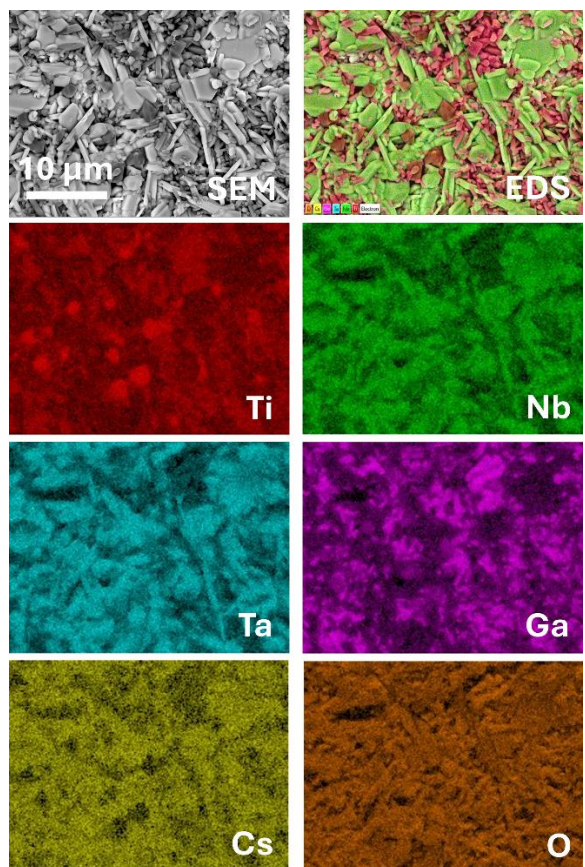

Fig. 3. Bimodal dispersion of elements in high-entropy oxide. SEM micrograph (top left), corresponding overlayed image of EDS maps (top right) and EDS elemental mappings with different elements for TiNbTaGaCsO$_9$.

Fig. 4 shows the oxidation state of TiNbTaGaCsO$_9$ examined by XPS. The peaks were deconvoluted to examine the correct peak position of each element. The presence of Ti$^{4+}$ is observed in Fig. 4(a) at 458.0 eV and 464.0 eV. The Nb$^{5+}$ peaks are also visible at 206.2 and 209.0 eV in Fig. 4(b), Ta$^{5+}$ peaks are seen at 25.3 and 27.0 eV in Fig. 4(c), Ga$^{3+}$ peaks are visible at 1116.7 and 1143.5 eV in Fig. 4(d), and Cs$^+$ peaks are visible at 723.7 and 737.7 eV in Fig. 4(e). These results confirm that all cations are in a fully oxidized state in TiNbTaGaCsO$_9$. Finally, the peak belonging to oxygen is observed at 529.2 eV in Fig. 4(d). The shift of the oxygen binding energy compared to conventional oxides [43] should be due to the chemical environment in this HEO [44]. In addition to the major peak of O$^{2-}$, a small peak towards high binding energy is detected, which is commonly attributed to the existence of hydroxyl and/or oxygen vacancies in different oxides [33,45].

The presence of oxygen vacancies with unpaired electrons can also be confirmed in ESR data in Fig. 5, in which a dual peak with a *g* value of 2.006 is observed. The combustion-based ONH analyzer also suggests 7.5% oxygen deficiency, which is reasonably close to the prediction of XPS (10.2%). Despite these estimations, it should be noted that neither XPS, ESR nor OHN data can provide reliable quantification for vacancies because XPS only detects surface vacancies, ESR only detects vacancies with unpaired electrons, and OHN underestimates oxygen content in the presence of hard-to-combust refractory metals. The vacancy existence in HEOs is not unexpected, because such defects are necessary to accommodate multiple cations with diverse valences and various atomic sizes in a lattice.

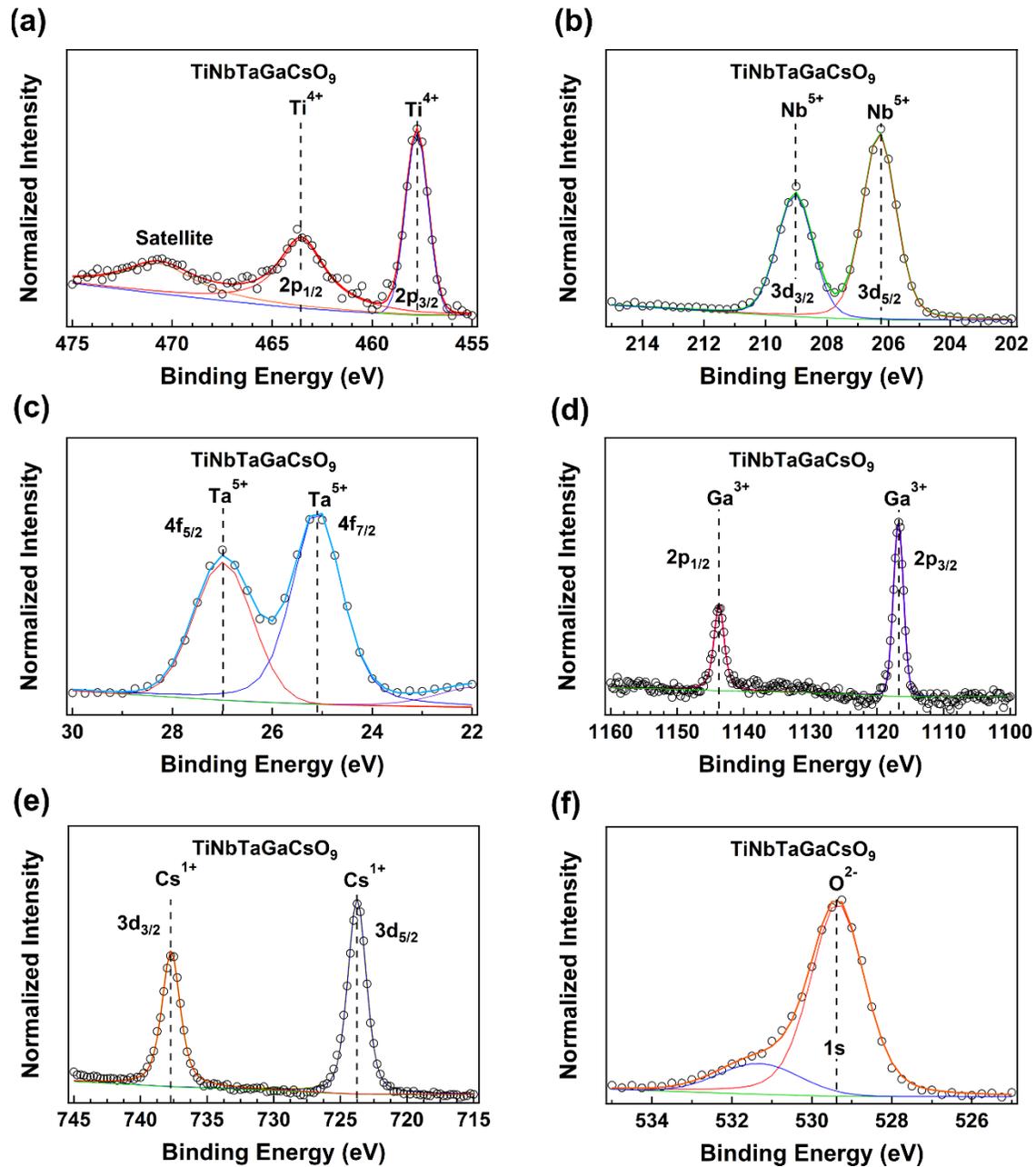

Fig. 4. Full oxidation states of cations in high-entropy oxide. XPS plots for (a) titanium, (b) niobium, (c) tantalum, (d) gallium, (e) cesium and (f) oxygen in TiNbTaGaCsO$_9$.

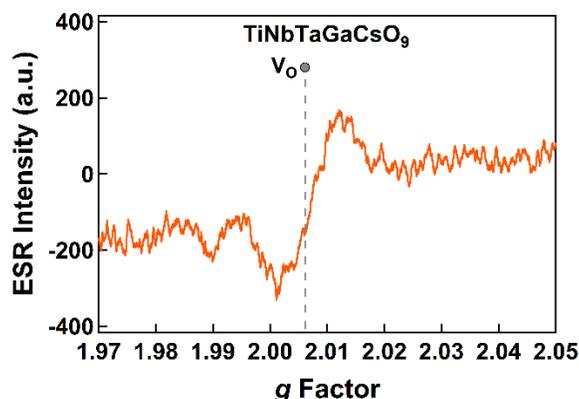

Fig. 5. Presence of oxygen vacancies in high-entropy oxide. ESR spectrum of TiNbTaGaCsO$_9$.

## 3.2. Optical properties

The optical properties of TiNbTaGaCsO$_9$ were analyzed by UV-vis spectroscopy and UPS, as illustrated in Fig. 6. TiNbTaGaCsO$_9$ is yellow in color, which indicates that it is able to absorb photons in the blue region of the light spectrum. According to the UV-vis spectrum, it has high light absorbance in both UV and visible light regions, although its light absorbance gradually decreases with the increment of the light wavelength, as observed in Fig. 6(a). The light absorbance pattern was evaluated using the Kubelka-Munk approach in Fig. 6(b), indicating that the bandgap of TiNbTaGaCsO$_9$ is 2.8 eV. The bandgap of the HEO is smaller than the bandgap of popular photocatalysts like TiO$_2$ (3.2 eV) [7], indicating promising optical properties of this HEO for photocatalysis. Fig. 6(c) shows the UPS plot at low binding energy. According to this spectrum, the location of the valence band maximum is at 1.6 ± 0.1 eV vs. NHE (normal hydrogen electrode). The combination of the information about the valence band and bandgap indicates that the band structure of the material is suitable for photocatalytic hydrogen generation and CO$_2$ photoconversion, as illustrated in Fig. 6(d). Fig. 6(d) shows that the energy requirements to perform different reactions (H$_2$O to O$_2$ at 1.23 eV vs. NHE, H$^+$ to H$_2$ at 0 eV vs. NHE [25], CO$_2$ to CO at -0.11 eV vs. NHE, and CO$_2$ to CH$_4$ at 0.17 eV vs. NHE [45,46]) lie within the bandgap.

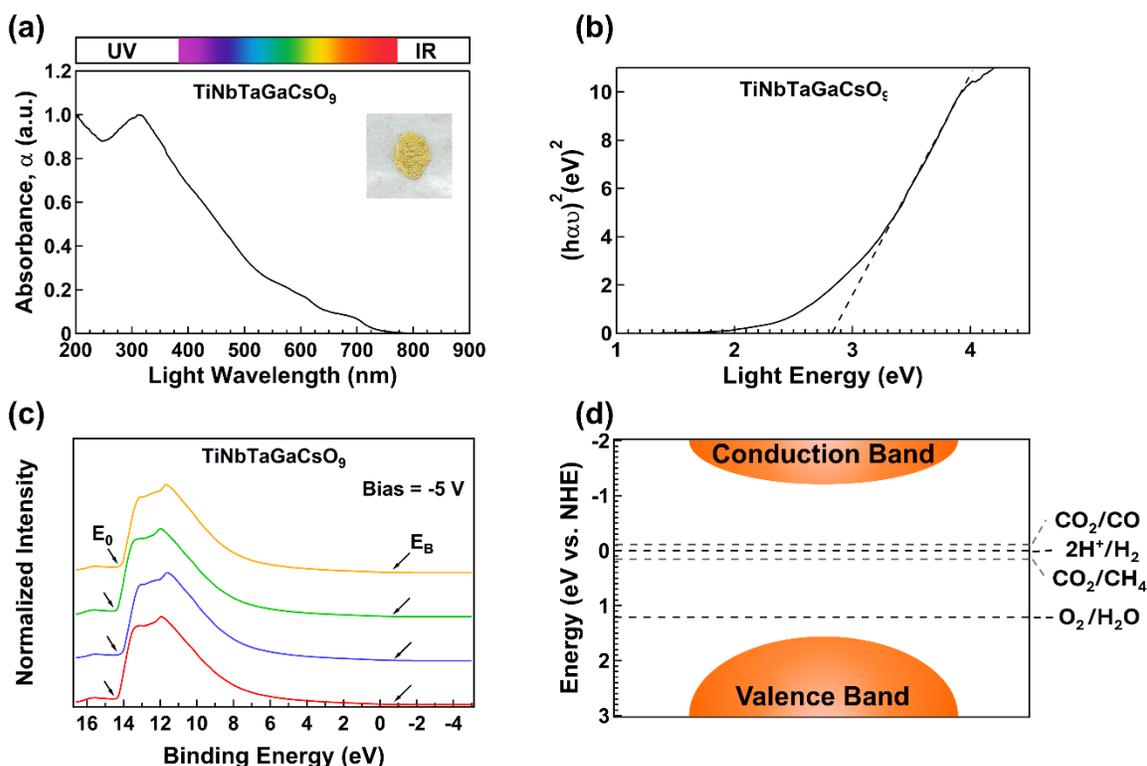

Fig. 6. Large light absorbance and proper band structure of high-entropy oxide for splitting $H_2O$ and converting $CO_2$. (a) Light absorbance achieved by UV-vis spectroscopy, (b) Kubelka-Munk plot, (c) valence band maximum determination by UPS at four different positions and (d) band structure of $TiNbTaGaCsO_9$. In (b), h and $\upsilon$ refer to Planck's constant and phonon frequency, respectively. In (c), $E_B$ is the valence band maximum versus the Fermi level, and $E_0$ is the secondary electron cutoff.

### 3.3. Photocatalytic tests

The results of (a) photocatalytic water splitting and (b-d) $CO_2$ conversion of $TiNbTaGaCsO_9$ are shown in Fig. 7. As mentioned in the methodology, the water decomposition test was carried out under three different conditions: (i) water, (ii) water and methanol, and (iii) water, methanol and platinum as a cocatalyst. As shown in Fig. 7(a), for the first water splitting condition without a sacrificial agent and a cocatalyst, hydrogen production only reaches 10.5 μmol per gram of the catalyst. Adding methanol consumes the photogenerated holes, allowing the photogenerated electrons to participate in the hydrogen evolution reactions, significantly increasing $H_2$ production to 0.8 mmol per gram of the catalyst. However, the highest activity was observed with the addition of a platinum cocatalyst, where the production reached 3.7 mmol per gram of the catalyst.

For $CO_2$ photoreduction, the results for CO, $CH_4$ and $H_2$ are shown in Fig. 7(b-d). The average CO, $CH_4$ and $H_2$ production rate for $TiNbTaGaCsO_9$ during 24 h of photocatalytic activity are 5.4, 1.1 and 42.4 μmol h$^{-1}$ per gram of the catalyst. Considering that the $CO_2$ photoconversion reaction and water splitting are competing in the reactor, the $CO_2$ reduction selectivity of this sample can be calculated using the following equation.

$CO_2$ Photoreduction Selectivity = $[8r_{CH_4} + 2r_{CO}] / [8r_{CH_4} + 2r_{CO} + 2r_{H_2}] \times 100$  (1)

Here, the production rate ($r$) is multiplied by the electron numbers required for different reactions: two electrons for CO, eight electrons for $CH_4$ and two electrons for $H_2$ formation [46,47]. The obtained selectivity for $CO_2$ reduction is calculated as 18.7%, suggesting that the sample exhibits a higher preference for hydrogen generation rather than $CO_2$ photoconversion. Similarly, the selectivity for methanation was also analyzed using the following equation [33], showing a selectivity of 45.2%, which is promising for methanation.

Selectivity for Methanation $= [8r_{CH4}] / [8r_{CH4} + 2r_{CO}] \times 100$ (2)

A higher selectivity for $H_2$ production compared to methanation should be due to a larger number of electrons that are needed for methanation: 8 against 2. Moreover, the production of $CH_4$ goes through some intermediate products, such as CO, and these intermediate products are usually detached from the surface of the catalyst without further contribution to methanation [14,15,28].

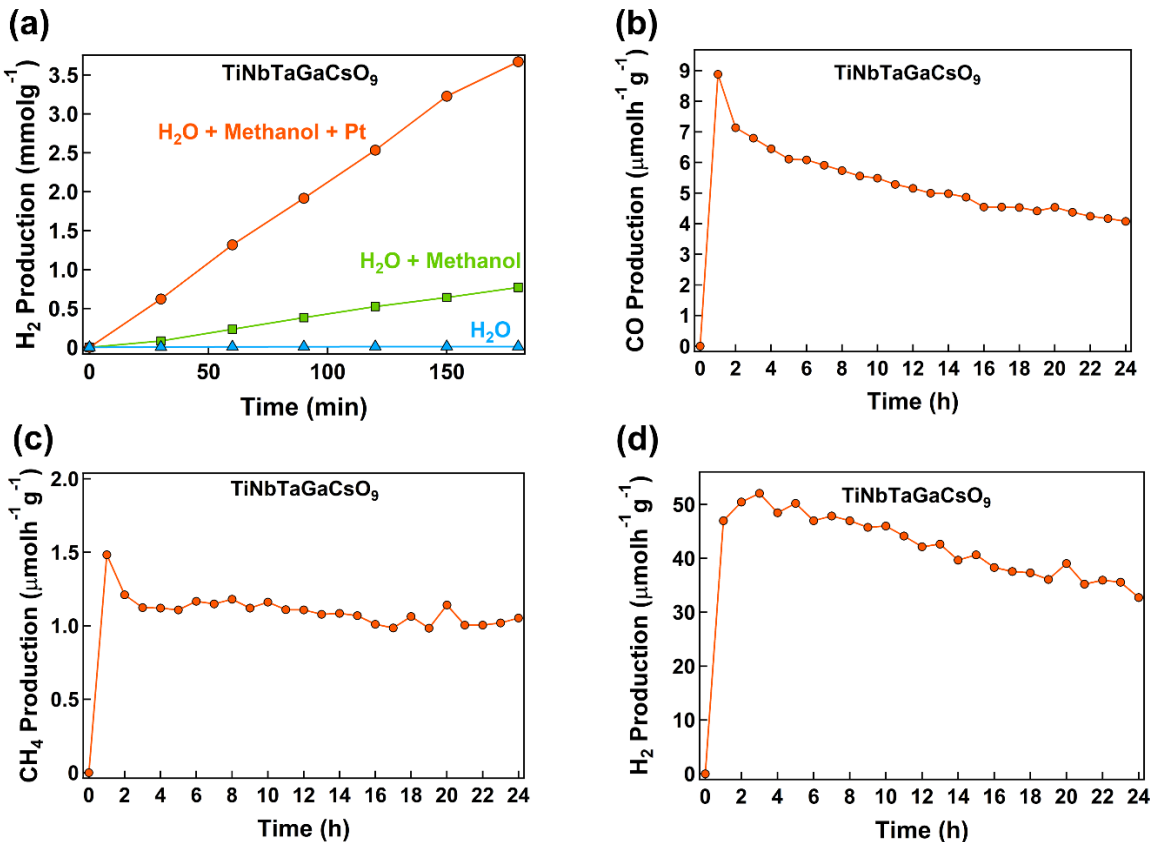

Fig. 7. Photocatalytic performance of high-entropy oxide for both hydrogen production and $CO_2$ conversion. (a) Hydrogen generation from water splitting by adding $TiNbTaGaCsO_9$ to three different solutions of (i) only water, (ii) water and methanol as sacrificial agent for holes and (iii) water and methanol plus platinum as cocatalyst. (b) CO, (c) $CH_4$ and (d) $H_2$ production from $CO_2$ photoreduction using $TiNbTaGaCsO_9$. Numbers are given per mass of catalyst.

Finally, the stability of the catalyst was examined first by a cycling test of water splitting reactions, and second by structural analysis using XRD after photocatalytic tests, as illustrated in Fig. 8. Fig. 8(a) displays the cycling test for $TiNbTaGaCsO_9$, indicating that the hydrogen production remains reasonably similar within 12 h at different cycles. Furthermore, the XRD patterns obtained after 12 h of water splitting and 24 h of $CO_2$ conversion test in Fig. 8(b) are consistent with the initial powder pattern. This indicates that the material does not suffer from any photodegradation or transformation due to the interaction with light or other chemicals utilized during the processes. The high stability is considered a typical characteristic of high-entropy materials [41,42].

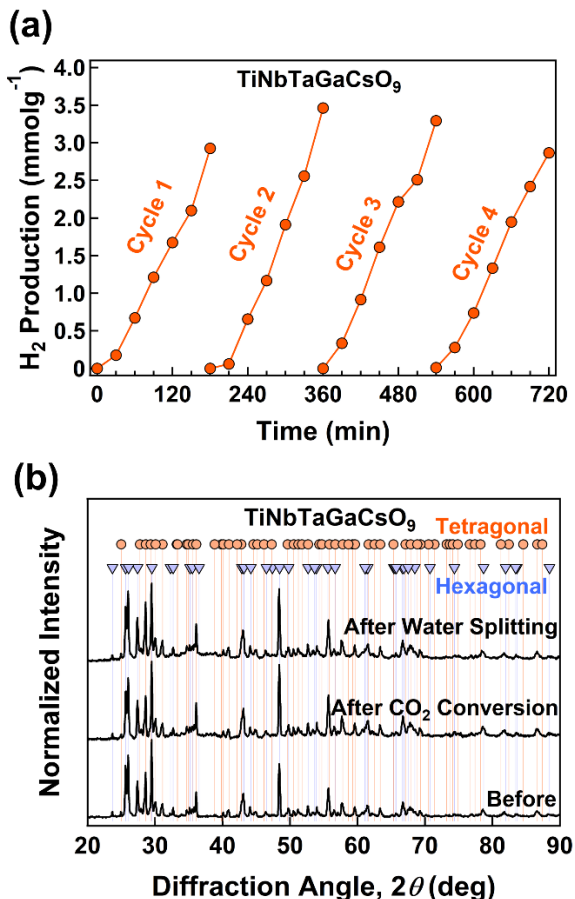

Fig. 8. High stability of high-entropy oxide photocatalyst. (a) Hydrogen production per catalyst mass versus irradiation time for four water splitting cycles using TiNbTaGaCsO$_9$. (b) Crystal structure of TiNbTaGaCsO$_9$ before and after photocatalytic hydrogen generation and CO$_2$ photoconversion.

## 4. Discussion

The study of high-entropy materials as photocatalysts provides numerous opportunities for innovating active catalysts. While previous studies on high-entropy materials have attributed their remarkable performance to a variety of factors, such as lattice strain [27,28], hybridized orbitals [25] and defects [28,30], the present work explores ways to design such catalysts based on a mixed $d^0+d^{10}+s^0$ cationic configuration strategy with a wide electronegativity mismatch. Building on this premise, a novel HEO, TiNbTaGaCsO$_9$, was developed. TiNbTaGaCsO$_9$ demonstrated high activity for hydrogen generation and CO$_2$ photoconversion. Here, two issues require detailed discussion: (i) assessment of the activity of the current HEO compared with other catalysts, and (ii) reasons for the high performance of this HEO.

Regarding the first issue, a comparison of the activity of TiNbTaGaCsO$_9$ ($d^0+d^{10}+s^0$ cationic configuration) with TiZrNbTaGaO$_{10.5}$ ($d^0+d^{10}$ cationic configuration) and TiZrHfNbTaO$_{11}$ (only $d^0$ cationic configuration) is shown in Fig. 9. As the specific surface area exhibits a clear effect on the performance of photocatalysts [35,45], the activity of materials is given per area (specific surface area of TiZrNbTaGaO$_{10.5}$ and TiZrHfNbTaO$_{11}$ are 0.48 m$^2$g$^{-1}$ and 0.75 m$^2$g$^{-1}$, respectively). TiNbTaGaCsO$_9$ demonstrates superior photocatalytic performance in comparison with TiZrNbTaGaO$_{10.5}$ and TiZrHfNbTaO$_{11}$, suggesting a benefit in the addition of alkali elements with the s$^0$ cationic configuration to the composition. Fig. 9 suggests that the catalytic photoactivity for hydrogen and methane formation improves in the order of d$^0$, d$^0$+d$^{10}$ and d$^0$+d$^{10}$+s$^0$, respectively (the lower CO production in the d$^0$+d$^{10}$ HEO compared to the d$^0$ HEO is due to the higher transformation of CO to CH$_4$). These features are not necessarily related to the band structure, as the bandgap of TiZrNbTaGaO$_{10.5}$ (2.5 eV) is smaller than that of TiNbTaGaCsO$_9$ (2.8 eV), while the bandgap of TiZrHfNbTaO$_{11}$ (3.2 eV) is larger than the two other oxides [25,30,34]. Therefore, it is crucial to highlight the difference between these materials and understand the effect of adding an alkali metal to the composition in the following paragraphs.

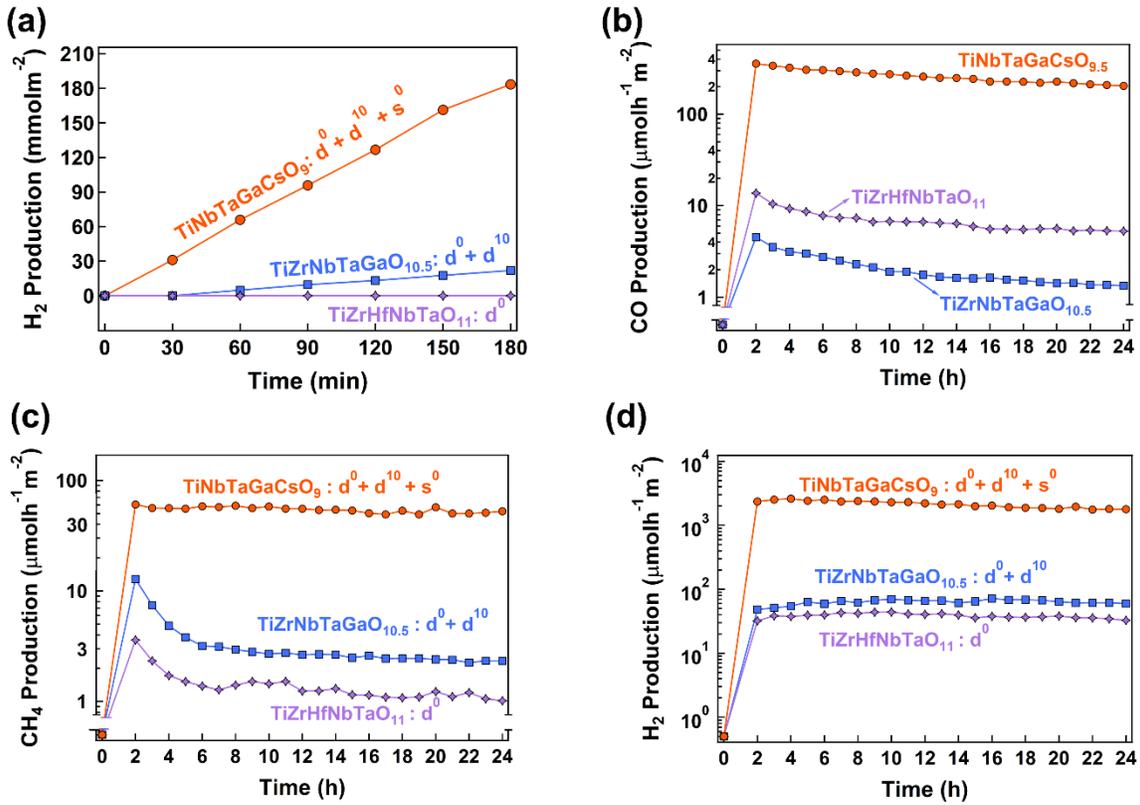

Fig. 9. Enhancement of photocatalytic activity by introducing wide electronegativity mismatch in high-entropy oxide with d$^0$+d$^{10}$+s$^0$ cationic configuration. (a) Hydrogen production from water splitting, (b) CO generation from CO$_2$ conversion, (c) CH$_4$ generation from CO$_2$ conversion and (d) H$_2$ generation from CO$_2$ conversion using three catalysts TiNbTaGaCsO$_9$ (d$^0$+d$^{10}$+s$^0$ configuration), TiZrNbTaGaO$_{10.5}$ (d$^0$+d$^{10}$ configuration) and TiZrHfNbTaO$_{11}$ (d$^0$ configuration). Numbers are given per surface area of catalysts.

Table 2 compares the photocatalytic activity of TiNbTaGaCsO$_9$ with some relevant published results in the literature [8,23,45,48–53]. The comparison between the data from different sources needs to be carefully done as the light source type, light intensity, concentration of catalyst and type of photoreactor can affect the photocatalytic reaction rates and even the apparent quantum yield [54]. Despite this limitation, Table 2 indicates that the HEO TiNbTaGaCsO$_9$ under identical light intensity exhibits a high hydrogen production and a high $CO_2$ conversion rate compared to other photocatalysts. $TiZrNbTaZnO_{10}$, which has a $d^0+d^{10}$ cationic structure, apparently shows a higher $CO_2$ conversion rate compared to $TiNbTaGaCsO_9$. However, the difference arises from the light intensity used in this study (230 mW cm$^{-2}$ for $TiNbTaGaCsO_9$) and in the literature (1400 mW cm$^{-2}$ for $TiZrNbTaZnO_{10}$) [53]. If the photocatalytic production rate of these two materials is normalized by light intensity, it can be concluded that CO and $CH_4$ production rates using $TiNbTaGaCsO_9$ are 2.1 and 1.1 times higher than $TiZrNbTaZnO_{10}$. Table 2 also shows that $TiNbTaGaCsO_9$ exhibits photocatalytic activities even better than state-of-the-art photocatalyst P25 $TiO_2$, confirming the significance of $d^0+d^{10}+s^0$ cationic configuration for the design of active high-entropy photocatalysts. It should be noted that in addition to high efficiency, high-entropy materials show good catalytic stability because of their structural stability resulting from a low Gibbs free energy [41,42].

Table 2. Comparison of several reported photocatalysts with high-entropy oxide $TiNbTaGaCsO_9$. Catalyst mass used for reaction, specific surface area, light source, $CO_2$ conversion rate and hydrogen production rate per photocatalyst surface area.

| Photocatalyst | Mass (mg) | Surface Area (m² g⁻¹) | Light Source | Light Intensity (mW cm⁻²) | $CO_2$ Conversion (μmol h⁻¹m²) | | | Water Splitting (mmol h⁻¹m⁻²) | Ref. |
|---|---|---|---|---|---|---|---|---|---|
| | | | | | CO | $CH_4$ | $H_2$ | $H_2$ | |
| Anatase $TiO_2$ | 50 | 13.50 | 300 W Xe | - | | | | 0.5 | [8] |
| P25 $TiO_2$ | 100 | 38.7 | 400W Hg | | 0.12 | | | | [30] |
| P25 $TiO_2$ | 200 | 52.7 | 300W Xe | | 0.02 | 0.05 | | | [48] |
| P25 $TiO_2$ | 5 | 50 | 100 W UV | 6.5 | | | | <0.1 | [49] |
| P25/Pd-Ba | 5 | 50 | 100 W UV | 6.5 | | | | 0.6 | [49] |
| $ZnGa_2O_4$ | 50 | 157.00 | 300 W Xe | - | | | | <0.1 | [50] |
| Cs-dopped g-$C_3N_4$ | 40 | 9.00 | Xe | 250 | | | | 0.4 | [23] |
| $CsPbBr_3$-NC/Amorphous-$TiO_2$ | 5 | | 150 W Xe | 150 | 28.2 | | | | [51] |
| $(Ga_{0.2}Cr_{0.2}Mn_{0.2}Ni_{0.2}Zn_{0.2})_3O_4$ | 20 | 16.71 | 300 W Xe | - | 1.4 | 0.2 | | | [44] |
| $(CdZnCuCoFe)S_{1.25}/ZnIn_2S_4$ | 10 | 62.20 | 300 W Xe | 119 | <0.1 | | | | [52] |
| $TiZrNbTaZnO_{10}$ | 100 | 0.03 | 400W Hg | 1400 | 761.3 | 301.0 | | | [53] |
| $TiZrNbTaZnO_{10}$ | 50 | 0.03 | 300 W Xe | 1500 | | | | 6.6 | [53] |
| $TiZrHfNbTaO_{11}$ | 100 | 0.66 | 400 W Hg | 230 | 7.4 | 1.6 | 36.4 | | This study |
| $TiZrHfNbTaO_{11}$ | 50 | 0.66 | 300 W Xe | 1400 | | | | <0.1 | This study |
| $TiZrNbTaGaO_{10.5}$ | 100 | 0.48 | 400 W Hg | 230 | 2.1 | 3.4 | 62.9 | | This study |
| $TiZrNbTaGaO_{10.5}$ | 50 | 0.48 | 300 W Xe | 1400 | | | | 7.8 | This study |
| $TiNbTaGaCsO_9$ | 100 | 0.03 | 400 W Hg | 230 | 268.4 | 55.3 | 2122.0 | | This study |
| $TiNbTaGaCsO_9$ | 50 | 0.03 | 300 W Xe | 1400 | | | | 62.9 | This study |

Regarding the second issue, previous studies highlighted that the co-presence of elements with mixed $d^0$ and $d^{10}$ cationic configurations improves the charge transfer of HEOs, by providing both good electron donor and acceptor sites [33,34,53]. First-principles calculations suggested that the elements with a lower electronegativity function as active sites for reactant adsorption, while elements with high electronegativity enhance the charge transfer [36]. Therefore, having a wide range of electronegativity by adding alkali metals with $s^0$ cationic configuration (e.g. cesium) and elements with $d^{10}$ cationic configuration (e.g. gallium) to conventional $d^0$ transition metal cations (e.g. titanium, niobium and tantalum) provides catalysts with both high reactant absorbance and good charge mobility. In addition to the electronegativity mismatch effects, individual cationic configurations play an essential role in improving the activity. For example, the presence of alkali metals in $d^0$ and $d^{10}$ metal oxides provokes lattice distortions and asymmetrical metal oxide octahedra, which generate polarization fields that enhance the ionic properties and water splitting activity [6]. Cesium enhances the position of the band structure, improves the reactant absorbance [23,55,56], suppresses the charge recombination and raises the structural stability [20,56]. These features of cesium have been used to develop some catalysts, such as $CsTaO_3$ [18,57] and $CsNbO_3$ [57]. Titanium, niobium and tantalum are important because the valence band is predominantly generated by O p orbitals and the conduction band by d orbitals of transition metals [36], and thus, $d^0$ cations usually contribute more significantly to the photocatalysis. It was also reported that the presence of gallium as a dopant can diminish charge recombination [34,44], while niobium and tantalum can improve charge separation [36]. All these features, together with a large fraction of oxygen vacancies as active sites, resulting from a large electronegativity mismatch, contribute to the high photocatalytic activity in this HEO.

Finally, it should be noted that the results of the photocatalytic tests with and without platinum for $TiNbTaGaCsO_9$ indicate the potential of high-entropy photocatalysts with a $d^0+d^{10}+s^0$ configuration for being cocatalyst-free. However, the improvement in $TiNbTaGaCsO_9$ with the cocatalyst addition, which results from the function of platinum as an electron sink to suppress charge recombination, indicates that such HEOs still need modifications to fully avoid the use of precious cocatalysts. Such modifications can be achieved by using $d^{10}$ elements with higher electronegativities as electron sinks, while keeping $s^0$ cations with a low electronegativity as reactant adsorption sites. Making the electronegativity mismatch wider is also expected to provide a positive effect on charge transfer to eventually replace the cocatalyst. However, proving the effect of electronegativity mismatch on charge carrier transfer requires complementary future experiments such as X-ray absorption spectroscopy, in situ XPS, Mott-Schottky electrochemical test, surface photovoltage measurement and density functional theory calculations. In addition to attempts to make these HEOs fully cocatalyst-free, future studies should also explore new strategies to synthesize nanopowders of these materials with a large surface area, because the HPT method used in this study generates large particle sizes [38].

**5. Conclusion**

A novel high-entropy oxide, $TiNbTaGaCsO_9$, was produced by mixing cations with $d^0$ (transition metals titanium, niobium and tantalum), $d^{10}$ (gallium with high electronegativity) and $s^0$ (alkali metal cesium with a low electronegativity) configurations. The photocatalyst exhibits enhanced performance for both hydrogen generation and carbon dioxide photoconversion. This study suggests that a mixed $d^0+d^{10}+s^0$ cationic configuration with wide electronegativity differences can be used as an effective strategy to design new photocatalysts with a large oxygen vacancy concenteration. Such a strategy, which was easily realized in a high-entropy material (materials with high tunability for compositional changes), can be basically employed to design any other photocatalysts.

**Declaration of Competing Interest**

The authors declare no competing financial interests or personal relationships that could influence the work reported in the current manuscript.

**Acknowledgment**

The author JHJ is grateful of the Q-Energy Innovator Fellowship of Kyushu University because of a scholarship. This work was supported partly by Mitsui Chemicals, Inc., Japan, and partly by the Japan Science and Technology Agency (No. JPMJFS2132 and No. JPMJAP2332).

**Data availability**
Data will be made available on request.